\documentclass[twocolumn]{aastex63}
\usepackage[caption=false]{subfig}
\usepackage{multirow}
\usepackage{amsmath}
\allowdisplaybreaks[1]

\received{}%\today}
\revised{}%\today}
\accepted{}%\today}
\submitjournal{ApJL}

\shorttitle{Evidence for Pinched Magnetic Fields in Quiescent Filaments of NGC~1333}
\shortauthors{Doi {\em et al.}}
\graphicspath{{./}{figures/}}
\begin{document}

\title{The JCMT BISTRO Survey: Evidence for Pinched Magnetic Fields in Quiescent Filaments of NGC~1333}

\correspondingauthor{Yasuo Doi}
\email{doi@ea.c.u-tokyo.ac.jp}

\author[0000-0001-8746-6548]{Yasuo~Doi}
\affiliation{Department of Earth Science and Astronomy, Graduate School of Arts and Sciences, The University of Tokyo, 3-8-1 Komaba, Meguro, Tokyo 153-8902, Japan}

\author[0000-0003-2726-0892]{Kohji Tomisaka}
\affiliation{National Astronomical Observatory of Japan, National Institutes of Natural Sciences, Osawa, Mitaka, Tokyo 181-8588, Japan}

\author[0000-0003-1853-0184]{Tetsuo Hasegawa}
\affiliation{National Astronomical Observatory of Japan, National Institutes of Natural Sciences, Osawa, Mitaka, Tokyo 181-8588, Japan}

\author[0000-0002-0859-0805]{Simon Coud\'e}
\affiliation{SOFIA Science Center, Universities Space Research Association, NASA Ames Research Center, M.S. N232-12, Moffett Field, CA 94035, USA}
\affiliation{Centre de Recherche en Astrophysique du Qu\'ebec (CRAQ), Universit\'e de Montr\'eal, D\'epartement de Physique, C.P. 6128 Succ. Centre-ville, Montr\'eal, QC H3C 3J7, Canada}

\author{Doris Arzoumanian}
\affiliation{Aix Marseille Univ, CNRS, CNES, LAM, Marseille, France}
\affiliation{National Astronomical Observatory of Japan, National Institutes of Natural Sciences, Osawa, Mitaka, Tokyo 181-8588, Japan}

\author[0000-0002-0794-3859]{Pierre Bastien}
\affiliation{Institut de Recherche sur les Exoplan\`etes (iREx), Universit\'e de Montr\'eal, D\'epartement de Physique, C.P. 6128 Succ. Centre-ville, Montr\'eal, QC H3C 3J7, Canada}
\affiliation{Centre de Recherche en Astrophysique du Qu\'ebec (CRAQ), Universit\'e de Montr\'eal, D\'epartement de Physique, C.P. 6128 Succ. Centre-ville, Montr\'eal, QC H3C 3J7, Canada}

\author[0000-0002-6906-0103]{Masafumi Matsumura}
\affiliation{Faculty of Education \& Center for Educational Development and Support, Kagawa University, Saiwai-cho 1-1, Takamatsu, Kagawa, 760-8522, Japan}

\author[0000-0001-8749-1436]{Mehrnoosh Tahani}
\affiliation{Dominion Radio Astrophysical Observatory, Herzberg Astronomy and Astrophysics Research Centre, National Research Council Canada, P. O. Box 248, Penticton, BC V2A 6J9, Canada}

\author[0000-0001-7474-6874]{Sarah Sadavoy}
\affiliation{Department for Physics, Engineering Physics and Astrophysics, Queen's University, Kingston, ON K7L 3N6, Canada}

\author[0000-0002-8975-7573]{Charles L. H. Hull}
\altaffiliation{NAOJ Fellow}
\affiliation{National Astronomical Observatory of Japan,
      Alonso de C\'{o}rdova 3788,
      Office 61B,
      7630422,
      Vitacura,
      Santiago,
      Chile}
\affiliation{Joint ALMA Observatory,
      Alonso de C\'{o}rdova 3107,
      Vitacura,
      Santiago,
      Chile}

\author[0000-0002-6773-459X]{Doug Johnstone}
\affiliation{Herzberg Astronomy and Astrophysics Research Centre, National Research Council of Canada, 5071 West Saanich Road, Victoria, BC V9E 2E7, Canada}
\affiliation{Department of Physics and Astronomy, University of Victoria, Victoria, BC V8P 5C2, Canada}

\author[0000-0002-9289-2450]{James Di Francesco}
\affiliation{Department of Physics and Astronomy, University of Victoria, Victoria, BC V8P 5C2, Canada}
\affiliation{Herzberg Astronomy and Astrophysics Research Centre, National Research Council of Canada, 5071 West Saanich Road, Victoria, BC V9E 2E7, Canada}

\author[0000-0001-9368-3143]{Yoshito Shimajiri}
\affiliation{Department of Physics and Astronomy, Graduate School of Science and Engineering, Kagoshima University, 1-21-35 Korimoto, Kagoshima, Kagoshima 890-0065, Japan}
\affiliation{National Astronomical Observatory of Japan, National Institutes of Natural Sciences, Osawa, Mitaka, Tokyo 181-8588, Japan}

\author[0000-0003-0646-8782]{Ray S. Furuya}
\affiliation{Institute of Liberal Arts and Sciences, Tokushima University, Minami Jousanajima-machi 1-1, Tokushima 770-8502, Japan}

\author[0000-0003-2815-7774]{Jungmi Kwon}
\affiliation{Department of Astronomy, Graduate School of Science, The University of Tokyo, 7-3-1 Hongo, Bunkyo-ku, Tokyo 113-0033, Japan}

\author[0000-0002-6510-0681]{Motohide Tamura}
\affiliation{Department of Astronomy, Graduate School of Science, The University of Tokyo, 7-3-1 Hongo, Bunkyo-ku, Tokyo 113-0033, Japan}
\affiliation{Astrobiology Center, National Institutes of Natural Sciences, Osawa, Mitaka, Tokyo 181-8588, Japan}
\affiliation{National Astronomical Observatory of Japan, National Institutes of Natural Sciences, Osawa, Mitaka, Tokyo 181-8588, Japan}

\author[0000-0003-1140-2761]{Derek Ward-Thompson}
\affiliation{Jeremiah Horrocks Institute, University of Central Lancashire, Preston PR1 2HE, UK}

\author[0000-0002-5714-799X]{Valentin J. M. Le Gouellec}
\affiliation{European Southern Observatory, Alonso de C\'{o}rdova 3107, Vitacura, Casilla 19001, Santiago, Chile}
\affiliation{Universit\'{e} Paris-Saclay, CNRS, CEA, Astrophysique, Instrumentation et Mod\'{e}lisation de Paris-Saclay, F-91191 Gif-sur-Yvette, France}

\author[0000-0003-2017-0982]{Thiem Hoang}
\affiliation{Korea Astronomy and Space Science Institute (KASI), 776 Daedeokdae-ro, Yuseong-gu, Daejeon 34055, Republic of Korea}
\affiliation{University of Science and Technology, Korea, 217 Gajang-ro, Yuseong-gu, Daejeon 34113, Republic of Korea}

\author[0000-0002-3036-0184]{Florian Kirchschlager}
\affiliation{Department of Physics and Astronomy, University College London, WC1E 6BT London, UK}

\author[0000-0001-7866-2686]{Jihye Hwang}
\affiliation{Korea Astronomy and Space Science Institute (KASI), 776 Daedeokdae-ro, Yuseong-gu, Daejeon 34055, Republic of Korea}
\affiliation{University of Science and Technology, Korea, 217 Gajang-ro, Yuseong-gu, Daejeon 34113, Republic of Korea}

\author[0000-0003-4761-6139]{Chakali Eswaraiah}
\affiliation{Indian Institute of Science Education and Research (IISER) Tirupati, Rami Reddy Nagar, Karakambadi Road, Mangalam (P.O.), Tirupati 517 507, India}

\author[0000-0003-2777-5861]{Patrick M. Koch}
\affiliation{Academia Sinica Institute of Astronomy and Astrophysics, No. 1, Sec. 4., Roosevelt Road, Taipei 10617, Taiwan}

\author[0000-0002-1178-5486]{Anthony P. Whitworth}
\affiliation{School of Physics and Astronomy, Cardiff University, The Parade, Cardiff, CF24 3AA, UK}

\author[0000-0002-8557-3582]{Kate Pattle}
\affiliation{Department of Physics and Astronomy, University College London, WC1E 6BT London, UK}

\author[0000-0003-4022-4132]{Woojin Kwon}
\affiliation{Department of Earth Science Education, Seoul National University, 1 Gwanak-ro, Gwanak-gu, Seoul 08826, Republic of Korea}
\affiliation{SNU Astronomy Research Center, Seoul National University, 1 Gwanak-ro, Gwanak-gu, Seoul 08826, Republic of Korea}

\author{Jihyun Kang}
\affiliation{Korea Astronomy and Space Science Institute (KASI), 776 Daedeokdae-ro, Yuseong-gu, Daejeon 34055, Republic of Korea}

\author[0000-0003-4366-6518]{Shu-ichiro Inutsuka}
\affiliation{Department of Physics, Graduate School of Science, Nagoya University, Furo-cho, Chikusa-ku, Nagoya 464-8602, Japan}

\author[0000-0001-7491-0048]{Tyler L. Bourke}
\affiliation{SKA Observatory, Jodrell Bank, Lower Withington, Macclesfield SK11 9FT, UK}
\affiliation{Jodrell Bank Centre for Astrophysics, School of Physics and Astronomy, University of Manchester, Oxford Road, Manchester, M13 9PL, UK}

\author[0000-0002-4154-4309]{Xindi Tang}
\affiliation{Xinjiang Astronomical Observatory, Chinese Academy of Sciences, 830011 Urumqi, People's Republic of China}

\author[0000-0001-9930-9240]{Lapo Fanciullo}
\affiliation{Academia Sinica Institute of Astronomy and Astrophysics, No. 1, Sec. 4., Roosevelt Road, Taipei 10617, Taiwan}

\author[0000-0002-3179-6334]{Chang~Won Lee}
\affiliation{Korea Astronomy and Space Science Institute (KASI), 776 Daedeokdae-ro, Yuseong-gu, Daejeon 34055, Republic of Korea}
\affiliation{University of Science and Technology, Korea, 217 Gajang-ro, Yuseong-gu, Daejeon 34113, Republic of Korea}

\author[0000-0003-3343-9645]{Hong-Li Liu}
\affiliation{Department of Astronomy, Yunnan University, Kunming, 650091, People's Republic of China}

\author[0000-0002-9907-8427]{A-Ran Lyo}
\affiliation{Korea Astronomy and Space Science Institute (KASI), 776 Daedeokdae-ro, Yuseong-gu, Daejeon 34055, Republic of Korea}

\author[0000-0002-5093-5088]{Keping Qiu}
\affiliation{School of Astronomy and Space Science, Nanjing University, 163 Xianlin Avenue, Nanjing 210023, People's Republic of China}
\affiliation{Key Laboratory of Modern Astronomy and Astrophysics (Nanjing University), Ministry of Education, Nanjing 210023, People's Republic of China}

\author[0000-0001-5522-486X]{Shih-Ping Lai}
\affiliation{Institute of Astronomy and Department of Physics, National Tsing Hua University, Hsinchu 30013, Taiwan}
\affiliation{Academia Sinica Institute of Astronomy and Astrophysics, No. 1, Sec. 4., Roosevelt Road, Taipei 10617, Taiwan}

\

%% Note that the \and command from previous versions of AASTeX is now
%% depreciated in this version as it is no longer necessary. AASTeX 
%% automatically takes care of all commas and "and"s between authors names.

%% AASTeX 6.3 has the new \collaboration and \nocollaboration commands to
%% provide the collaboration status of a group of authors. These commands 
%% can be used either before or after the list of corresponding authors. The
%% argument for \collaboration is the collaboration identifier. Authors are
%% encouraged to surround collaboration identifiers with ()s. The 
%% \nocollaboration command takes no argument and exists to indicate that
%% the nearby authors are not part of surrounding collaborations.

%% Mark off the abstract in the ``abstract'' environment. 
\begin{abstract}
We investigate the internal 3D magnetic structure of dense interstellar filaments within NGC 1333 using polarization data at $850~\mu\mathrm{m}$ from the \emph{B}-fields In STar-forming Region Observations survey at the James Clerk Maxwell Telescope.
Theoretical models predict that the magnetic field lines in a filament will tend to be dragged radially inward (i.e., pinched) toward the central axis due to the filament's self-gravity.
We study the cross-sectional profiles of the total intensity (\emph{I}) and polarized intensity (PI) of dust emission in four segments of filaments unaffected by local star formation that are expected to retain a pristine magnetic field structure.
We find that the filaments' FWHMs in PI are not the same as those in \emph{I}, with two segments being appreciably narrower in PI (FWHM ratio $\simeq 0.7$--0.8) and one segment being wider (FWHM ratio $\simeq 1.3$).
The filament profiles of the polarization fraction (\emph{P}) do not show a minimum at the spine of the filament, which is not in line with an anticorrelation between \emph{P} and \emph{I} normally seen in molecular clouds and protostellar cores.
Dust grain alignment variation with density cannot reproduce the observed \emph{P} distribution.
We demonstrate numerically that the \emph{I} and PI cross-sectional profiles of filaments in magnetohydrostatic equilibrium will have differing relative widths depending on the viewing angle.
The observed variations of FWHM ratios in NGC 1333 are therefore consistent with models of pinched magnetic field structures inside filaments, and especially if they are magnetically near-critical or supercritical.
\end{abstract}

%% Keywords should appear after the \end{abstract} command. 
%% See the online documentation for the full list of available subject
%% keywords and the rules for their use.
\keywords{
Interstellar medium (847);
Interstellar magnetic fields (845);
Interstellar filaments (842);
Star formation (1569);
Polarimetry (1278); Submillimeter astronomy (1647)
}
        
%% From the front matter, we move on to the body of the paper.
%% Sections are demarcated by \section and \subsection, respectively.
%% Observe the use of the LaTeX \label
%% command after the \subsection to give a symbolic KEY to the
%% subsection for cross-referencing in a \ref command.
%% You can use LaTeX's \ref and \label commands to keep track of
%% cross-references to sections, equations, tables, and figures.
%% That way, if you change the order of any elements, LaTeX will
%% automatically renumber them.
%%
%% We recommend that authors also use the natbib \citep
%% and \citet commands to identify citations.  The citations are
%% tied to the reference list via symbolic KEYs. The KEY corresponds
%% to the KEY in the \bibitem in the reference list below. 

\section{Introduction}
\label{sec: introduction}

It is widely recognized that filaments in the interstellar medium (ISM) play an essential role in the star formation process \citep[e.g.,][]{2014prpl.conf...27A}.
Theoretical studies indicate that the magnetic field (\emph{B}-field hereafter) contributes to the evolution of these filaments \citep[e.g.,][]{2019FrASS...6....5H}.
It is, therefore, crucial to observe the \emph{B}-field in quiescent filaments before the onset of star formation to understand their dynamical importance in shaping these ubiquitous structures.

Specifically, the plane-of-sky (POS) component of the \emph{B}-field can be traced with polarimetric observations of the thermal continuum emission from interstellar dust particles \citep[e.g.,][]{1988QJRAS..29..327H}.
Aspherical dust particles irradiated by incoming radiation fields are spun up by radiative alignment torques (RATs; \citealp{2007MNRAS.378..910L}), which align their rotation axes parallel to the ambient \emph{B}-field direction.
As a result, the thermal emission from so-aligned dust particles is polarized, and the polarization angle is perpendicular to the POS-projected \emph{B}-field \citep{1966ApJ...144..318S,1988QJRAS..29..327H}.

For a uniform \emph{B}-field along the line of sight (LOS), the polarized intensity (PI) and, similarly, the polarization fraction (\emph{P}) relative to the total intensity (\emph{I}) depend on the degree of alignment of the dust particles.
Assuming that the alignment is produced by the surrounding radiation field (i.e., RAT theory), \emph{P} will become smaller in high gas density regions shielded from this radiation \citep{2021ApJ...908..218H}.

Also, \emph{P} has a dependence on the viewing angle of the \emph{B}-field.
If the \emph{B}-field is highly inclined relative to the POS, \emph{P} can be lower, since the rotation axes of aspherical dust particles become nearly parallel to the LOS in such arrangements.
Moreover, the \emph{B}-field itself can be complicated within the observational beam or along the LOS by, e.g., gas turbulent motions.
Unresolved polarization structures will result in depolarization, as multiple position angles within the telescope beam cancel out and reduce the observed \emph{P} (geometric depolarization).

\defcitealias{2020ApJ...899...28D}{Paper~I}
In the Perseus molecular cloud, NGC 1333 is an active star-forming region with a complex network of massive (gravitationally supercritical) filaments (e.g., \citealp{2017A&A...606A.123H}).
\citet[][hereafter \citetalias{2020ApJ...899...28D}]{2020ApJ...899...28D} made a polarimetry study of NGC 1333 as a part of the \emph{B}-fields In STar-forming Region Observations (BISTRO) survey using the Sub-millimeter Common-User Bolometer Array 2 (SCUBA-2) camera and its polarimeter (POL-2) on the James Clerk Maxwell Telescope (JCMT).
The distance to this area is estimated to be $\simeq 300$ pc (\citealp{2018ApJ...869...83Z,2019ApJ...879..125Z,2021A&A...645A..55P}), giving a spatial resolution of 0.02 pc for the $14\arcsec.1$ JCMT beam (FWHM; \citealp{2013MNRAS.430.2534D}) at $850~\mu\mathrm{m}$.
With this high spatial resolution, \citetalias{2020ApJ...899...28D} spatially resolved the polarized emission from these filaments for the first time.
In this paper, we take advantage of these same data to investigate the 3D morphology of the \emph{B}-field inside several quiescent massive filaments.

This paper is organized as follows.
In Section \ref{sec:ObsReduce}, we outline our data reduction, with full details given in \citetalias{2020ApJ...899...28D}, together with our estimation of the cross-sectional profiles of filaments in \emph{I}, PI, and \emph{P}.
In Section \ref{sec: results}, we describe the characteristics of the estimated cross-sectional profiles.
In Section \ref{sec: model}, we compare our observations with a magnetohydrostatic simulation \citep{2014ApJ...785...24T} and investigate the 3D \emph{B}-field morphology inside the filament.
We discuss the assumptions and caveats of this work in Section \ref{sec:discussion}.
In Section \ref{ref:conclusions}, we conclude with a summary of our results.

\section{Observations and Methods}
\label{sec:ObsReduce}

\begin{figure*}[tp]
\centering
\includegraphics[width=\linewidth]{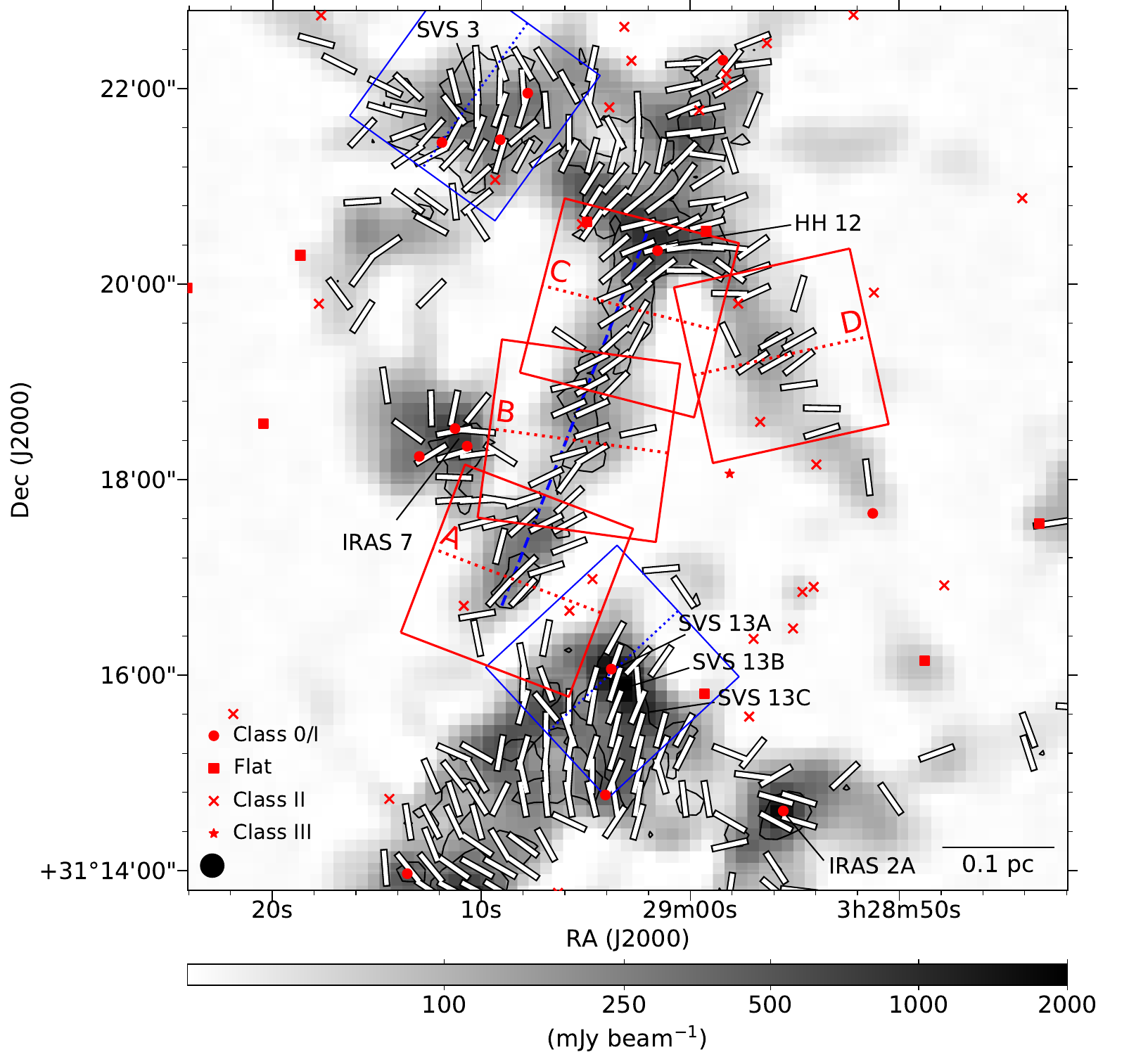}
\caption{
Four NGC 1333 positions where we measure cross sections of filaments. The red dotted lines are the positions of cross sections shown in Figure \ref{fig:filament_section_profile}.
The red boxes correspond to the smoothed maps we show in Figure \ref{fig:filament_section_map} (see text).
The blue dotted lines with blue boxes are the locations of the example filament profiles for the active regions for comparison, shown in Figures \ref{fig:filament_section_map_app} and \ref{fig:filament_section_profile_app}.
These profiles are briefly described in Section \ref{sec:discussion}.
The blue dashed line is the position where we estimate spatial power spectra of \emph{I} and PI to confirm that we observe small-scale structures in these emissions with equal sensitivities.
The gray scale is the $850~\mu\mathrm{m}$ Stokes $I$ map, and black contours indicate $\mathrm{PI}/\delta \mathrm{PI} \geqslant 5$ (\citetalias{2020ApJ...899...28D}).
White line segments indicate the POS magnetic field orientations (\citetalias{2020ApJ...899...28D}).
We plot the magnetic field orientations for the data points with $I \geqslant 25~\mathrm{mJy~beam^{-1}}$ and $\mathrm{PI} / \delta \mathrm{PI} \geqslant 3$.
Note that the length of the line segments has been normalized to show only the orientation of the magnetic field. 
The positions of the main infrared sources indicated in the figure are taken from \citet{2001ApJ...546L..49S}.
Red symbols are embedded YSOs ($A_V \geqslant 3$ mag; \citealp{2009ApJS..181..321E,2015AJ....150...40Y}).
Circles: class 0/I; squares: flat, crosses: class II, stars: class III.
The JCMT beam ($14\arcsec .1$) is shown in the lower left corner.
A reference scale for 0.1 pc is shown, for which we assume the distance to the source to be 300 pc (\citealp{2018ApJ...865...73O}; \citealp{2018ApJ...869...83Z}; see \citetalias{2020ApJ...899...28D}).
}
\label{fig:filament_section_pos}
\end{figure*}

We use the same polarimetry data as in \citetalias{2020ApJ...899...28D} (see that paper for full details of the observations and data analysis).
The observation covers the main part ($\sim 1.5 \mathrm{pc} \times 2.0 \mathrm{pc}$) of NGC 1333, and an intricate network of filaments with a column density above $\sim 10^{23}~\mathrm{H_2~cm^{-2}}$ is detected in \emph{I} and PI with a good signal-to-noise ratio (S/N) of $I/\delta I \geqslant 10$ and $\mathrm{PI}/\delta \mathrm{PI} \geqslant 3$.
A central cutout of the observation is shown in Figure \ref{fig:filament_section_pos}.
We check the \emph{I} and PI sensitivity as a function of the spatial frequency by estimating the spatial power spectra of \emph{I} and PI along the blue dashed line in Figure \ref{fig:filament_section_pos}.
We confirm that two observations trace the small-scale structure comparable to the filament width with equal sensitivity (Appendix \ref{sec:polynomial}).

In \citetalias{2020ApJ...899...28D}, we identified five filaments in the observed region.
Here we obtain cross-sectional profiles of these filaments for locations that satisfy each of the following conditions.
\begin{enumerate}
\item The S/N of the PI emission $\mathrm{PI}/\delta \mathrm{PI} \geqslant 3$ over more than three contiguous observational beams.
\item No star formation activity is found in its vicinity (see Figure \ref{fig:filament_section_pos} and Appendix \ref{sec:dust temperature}).
Thus, the radiation and dynamic interaction via, e.g., bipolar outflows from active star-forming regions that may affect the filament profile is expected to be negligible.
\end{enumerate}

We identify four filament segments, A, B, C, and D, in Figure \ref{fig:filament_section_pos} that fulfill these conditions.
All but segment D are in fact identified with $\mathrm{PI} /\delta \mathrm{PI} \geqslant 5$.
Segments B and C are bridged by radiation with $\mathrm{PI} /\delta \mathrm{PI} \geqslant 3$, but we treat them as two independent regions because of their clear separation, which can be seen in the contour with $\mathrm{PI} /\delta \mathrm{PI} = 5$ (Figure \ref{fig:filament_section_pos}).

We measure the Stokes \emph{I}, \emph{Q}, and \emph{U} intensities in each segment in the same way as in \citetalias{2020ApJ...899...28D}.
Additionally, we smooth the data in the direction along the segment five times the beam to boost the S/N by a factor of $\sqrt{5}$.
See Appendix \ref{sec:profiles} and Figure \ref{fig:filament_section_map} for the details of the derivation method and the resulting intensity maps.

The positions over which we estimate the filament cross-sectional profiles are shown as horizontal dotted lines in Figure \ref{fig:filament_section_pos}.
For segment A, the dotted line correspond to the peak in PI intensity.
For segment B, the PI profile of the peak position shows the contribution of another peak visible in the upper right part of the PI map (the southern extension of segment C; see Figures \ref{fig:filament_section_pos} and \ref{fig:filament_section_map}), resulting in a relatively large fitting error.
Hence, we choose a location one beam south of the PI peak to avoid the influence of this additional peak to achieve a Gaussian fit with reasonable accuracy.
The difference in fitting results between these two positions is not statistically significant, and this position shift does not affect the following discussion.
For segment C, the PI and \emph{I} intensities increase continuously toward HH12 in the north, so we set the position of C near the middle of the filament to also exclude a class 0/I young stellar object (YSO; J032901.56+312020.6; see Figure \ref{fig:filament_section_map}) from the evaluation of the cross section.
For segment D, we ignore one class II YSO (J032856.12+311908.4; see Figure \ref{fig:filament_section_map})
as a foreground source because it has an estimated $A_V = 0$ mag \citep{2009ApJS..181..321E}, and the source probably does not affect the filament's internal structure.
We thus set the position of D at the peak in PI intensity.

The projected distances to the neighboring YSOs from the centers of the cross sections are $\geqslant 0.05$ pc.
Stellar flux from a classical T Tauri star at 0.05 pc corresponds to $\sim0.03$--0.3 of the average interstellar radiation field \citep{2014ApJ...784..127F}.
The flux will be smaller if there is absorption between the star and the filament.
Thus, we judge that the radiation from the neighboring YSOs on the selected filament segments is negligible.
We further check the dust temperature distribution \citep{2021A&A...645A..55P} and find that both embedded heating source(s) and nonuniform external heating from nearby sources are negligible (Appendix \ref{sec:dust temperature}).
In addition, there is no signs of outflows from YSOs in these filament segments \citep{2020A&A...641A..36D}.
Also, we find no indication of interaction between outflows and filaments in the observed \emph{B}-field morphology (\citetalias{2020ApJ...899...28D}).

\section{Results}
\label{sec: results}

\begin{figure*}[tp]
\centering
\includegraphics[width=0.95 \linewidth]{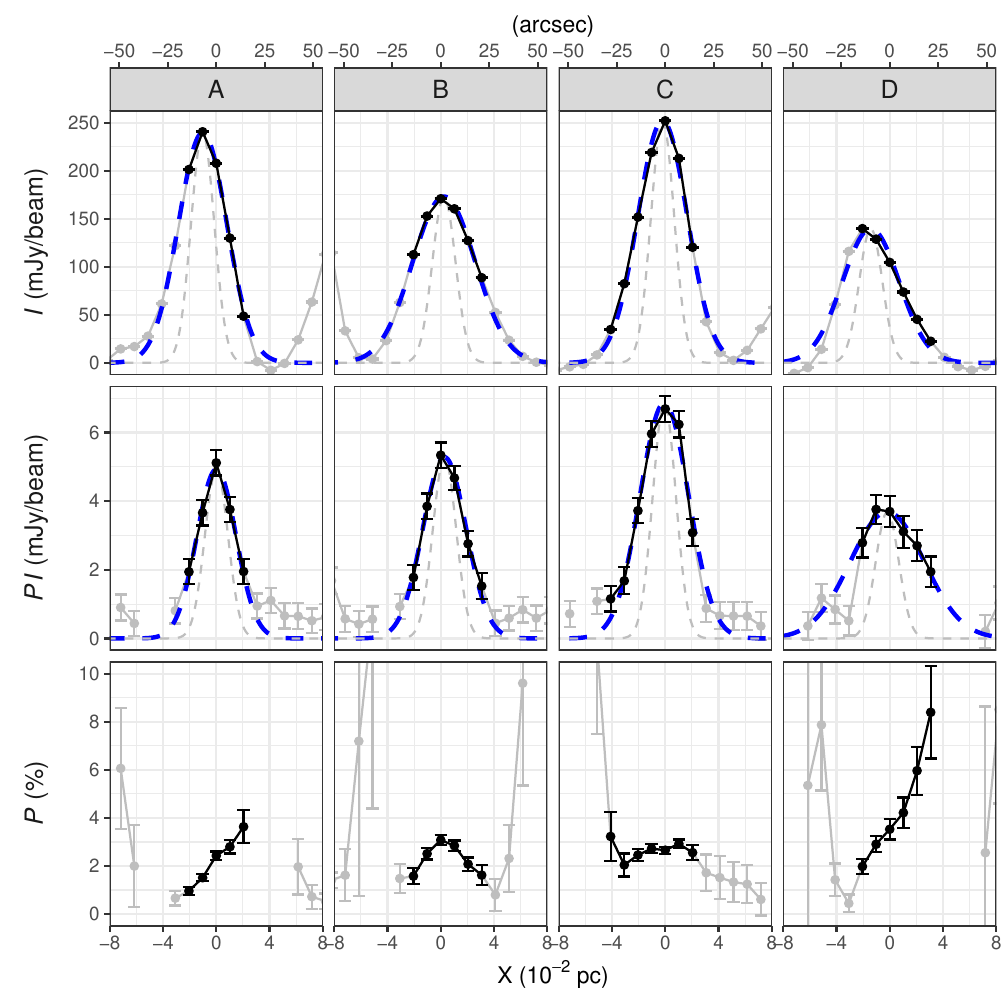}
\caption{Cross-sectional profiles of the selected NGC 1333 filamentary structures.
The X-axis points west, and the positions of the cross sections are indicated as red dotted lines in Figure \ref{fig:filament_section_pos}.
See Appendix \ref{sec:profiles} for the intensity and its error estimations.
We performed Gaussian fits using $\mathrm{PI}/\delta \mathrm{PI} \geq 3$ data, displayed in black in the cross-sectional profiles, and we do not use the data with $\mathrm{PI}/\delta \mathrm{PI} < 3$, shown in gray for the fitting (Appendix \ref{sec:profiles}).
The fitting results are shown as blue dashed lines in those panels.
The gray dashed lines are the beam profile of the observation ($\mathrm{FWHM}=14\arcsec .1$) for comparison.
Note that the data points in each cross section are plotted twice per beam, while fitting is performed using only one point per beam.
}
\label{fig:filament_section_profile}
\end{figure*}

We show the measured cross-sectional profiles of \emph{I}, PI, and \emph{P} of the filament segments in Figure \ref{fig:filament_section_profile}.
We estimate the FWHM of the \emph{I} and PI profiles with $\mathrm{PI}/\delta \mathrm{PI} \geqslant 3$ (black symbols in the figure) by fitting 1D Gaussian profiles (see Appendix \ref{sec:profiles} for the details), which we show as blue dashed lines.
The results are summarized in Table \ref{table:FWHM}.

\begin{table}[bt]
\caption{Estimated FWHM Values}
\label{table:FWHM}
\begin{center}
\begin{tabular}{lccc}
\hline
\hline
Segment & $\mathrm{FWHM}(I)$ & $\mathrm{FWHM}(\mathrm{PI})$ & \multirow{2}{*}{$\dfrac{\mathrm{FWHM}(\mathrm{PI})}{\mathrm{FWHM}(I)}$}\\
& (pc) & (pc)\\
\hline
A  & $0.043 \pm 0.001$ & $0.034 \pm 0.002$ & $0.81 \pm 0.04$\\
 & ($0.036 \pm 0.001$) & ($0.027 \pm 0.002$) & ($0.74 \pm 0.05$)\\
B & $0.055 \pm 0.002$ & $0.037 \pm 0.002$ & $0.69 \pm 0.04$\\
 & ($0.050 \pm 0.002$) & ($0.031 \pm 0.002$) & ($0.63 \pm 0.04$)\\
C & $0.044 \pm 0.002$ & $0.042 \pm 0.003$ & $0.98 \pm 0.07$\\
 & ($0.037 \pm 0.001$) & ($0.037 \pm 0.003$) & ($0.98 \pm 0.08$)\\
D & $0.050 \pm 0.003$ & $0.064 \pm 0.006$ & $1.31 \pm 0.15$\\
 & ($0.044 \pm 0.003$) & ($0.061 \pm 0.006$) & ($1.37 \pm 0.17$)\\
\hline
\multicolumn{4}{p{0.45\textwidth}}{{\bf Note.} We tabulate observed values without parentheses and beam-deconvolved values with parentheses.
See Appendix \ref{sec:profiles} for the details of the derivation.}
\end{tabular}
\end{center}
\end{table}

The estimated FWHM of the PI is appreciably narrower than that of $I ~(\mathrm{FWHM}(\mathrm{PI})/\mathrm{FWHM}(I) < 1)$ for segments A and B.
The differences from a ratio of 1.0 are statistically significant: $4.3 \sigma$ for segment A and $8.7 \sigma$ for segment B.
The differences become even more significant if we deconvolve the observations with the observational beam: $5.2 \sigma$ for segment A and $10.0 \sigma$ for segment B (the numbers with parentheses in Table \ref{table:FWHM}).
For segment C, $\mathrm{FWHM}(\mathrm{PI})/\mathrm{FWHM}(I) \simeq 1$.
For segment D, on the other hand, $\mathrm{FWHM}(\mathrm{PI})/\mathrm{FWHM}(I) > 1$ by $2.1 \sigma$, though the PI profile is not well fitted with a single Gaussian, partly due to the lower S/N of the PI signal.

As for the \emph{P} profiles, segment B shows an apparent, albeit slight, increase in \emph{P} right at its spine position.
This \emph{P} increase at the filament's spine is consistent with the filament's narrower PI profile compared  to that of \emph{I}.
On the other hand, this \emph{P} increase in the filament interior is not in line with the negative correlation between \emph{P} and \emph{I} normally found in the interior of dense clouds \citep[e.g.,][]{2021A&A...647A..78A}.
In segment C, where $\mathrm{FWHM}(\mathrm{PI})/\mathrm{FWHM}(I) \simeq 1$, the \emph{P} value is nearly constant within the filament.
In segments A and D, \emph{P} increases from one side of the filament to the other.
This is due to the offset of the filament center positions in \emph{I} relative to PI by $\simeq 0.01$ pc.
In summary, none of the \emph{P} profiles show the anticorrelation with \emph{I} normally found in molecular clouds and protostellar cores.

As a result, we find that the $\mathrm{FWHM}(\mathrm{PI}) / \mathrm{FWHM}(I)$ ratio varies between 0.7 and 1.3 depending on filament segment.
The relatively narrow and more centralized cross-sectional profile of PI to that of \emph{I} ($\mathrm{FWHM}(\mathrm{PI}) / \mathrm{FWHM}(I) < 1$) results in an increase of \emph{P} at the spine of the filament (e.g., the \emph{P} profile of segment B).
As described in Section \ref{sec: introduction}, the increase in \emph{P} is caused, for example, by a more efficient alignment of dust particles.
However, according to the RAT theory, increasing dust alignment inside dense filaments shielded from the ambient radiation field is challenging (Section \ref{sec: introduction}).
On the other hand, the observed variation of $\mathrm{FWHM}(\mathrm{PI}) / \mathrm{FWHM}(I)$ may be characterized by the LOS variation of the \emph{B}-field orientation.
The observed narrow and distinct PI profiles suggest that we need to consider the change of orientation angle of the POS-projected \emph{B}-field along the LOS within a filament.

In the following section, we discuss the possible cause of the variation of $\mathrm{FWHM}(\mathrm{PI}) / \mathrm{FWHM}(I)$.

\section{Pinched \emph{B}-field Predicted by the Tomisaka (2014) Model}
\label{sec: model}

\citetalias{2020ApJ...899...28D} observed that the \emph{B}-field aligns with different offset angles with respect to the major axis of each filament.
They attributed this distribution to observing mutually orthogonal \emph{B}-fields and filaments at different viewing angles.
The probability distribution of the offset angles between each pair of a POS \emph{B}-field and its associated filament is also consistent with this claim \citepalias{2020ApJ...899...28D}.

\begin{figure*}[tp]
\centering
\includegraphics[width=1.0\textwidth]{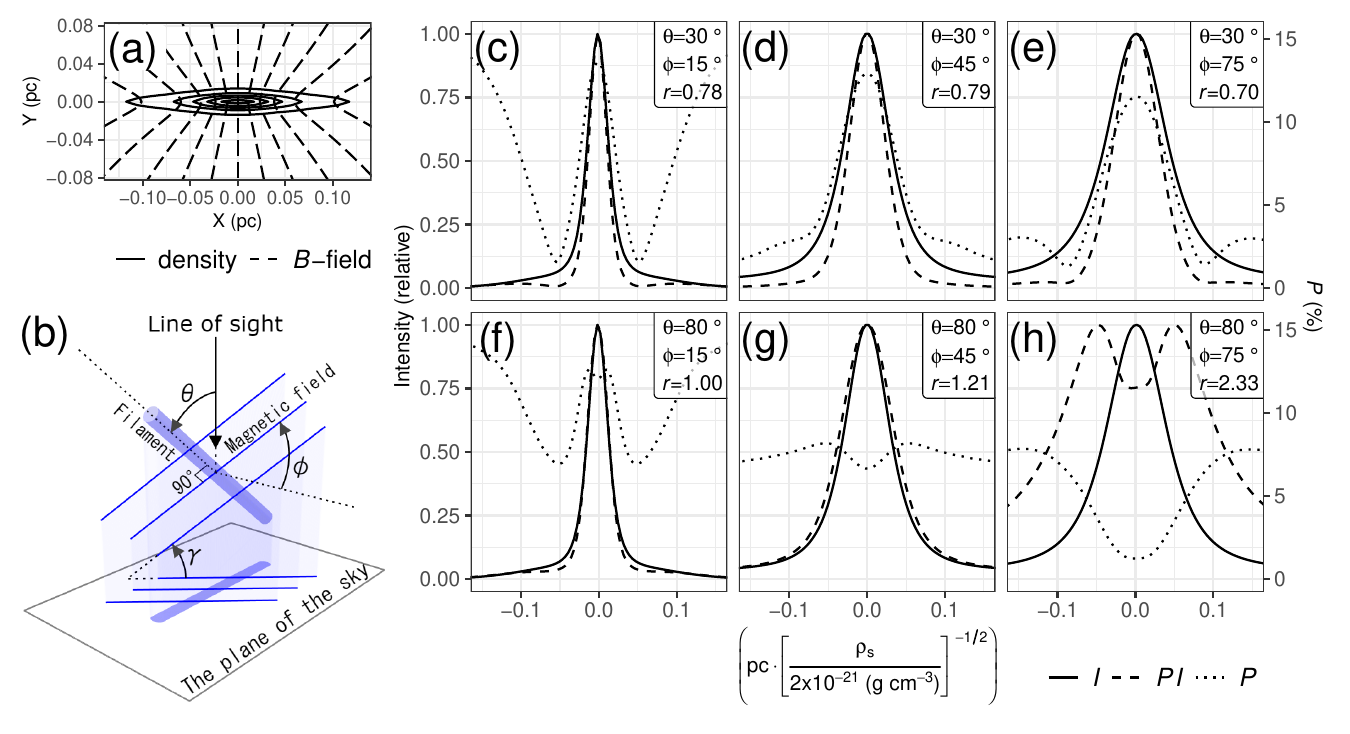}
\caption{
(a) Cross-sectional density profile of a filament perpendicular to the \emph{X}--\emph{Y} plane (solid contours) and its magnetic field line shapes (dashed lines), predicted by \citet{2014ApJ...785...24T}.
The density contour levels are $\rho/\rho_c = 0.1,~0.3,~\ldots,~0.9$.
The assumed model parameters are initial gas distribution radius $R_0=5$, thermal-to-magnetic pressure ratio $\beta_0=0.1$, and filament center-to-surface  density ratio $\rho_c/\rho_s=1000$ at equilibrium \citep{2014ApJ...785...24T}.
(b) Definitions of the relative angle of the filament with respect to the LOS ($\theta$), relative angle of the magnetic field to the POS ($\gamma$), and the rotation angle of the magnetic field with respect to the long axis of filament ($\phi$).
The origin of $\phi$ is defined as the angle when the magnetic field is on the POS.
(c)--(h) Predicted profiles of \emph{I} (solid line), PI (dashed line), and \emph{P}
(dotted line) when observing the filament shown in panel (a) from various directions of $\theta$ and $\phi$.
The \emph{r} in the upper right corner of each figure denotes the FWHM ratio of PI to \emph{I}.
We assume $2\times10^{-21}$ (g cm$^{-3}$) as the surface density of the filament when converting the radial distribution of the gas to a physical scale in parsecs.
}
\label{fig:plummer_profile}
\end{figure*}

\citet{2014ApJ...785...24T} studied the magnetohydrostatic equilibrium solution for an isothermal gas in a filament that is orthogonal to the \emph{B}-field (also see \citealp{2015ApJ...807...47T,2021ApJ...920..161T,2021ApJ...911..106K}).
His model (hereafter the Tomisaka model) parameters include the radius of the hypothetical parent cloud\footnote{The radius within which the filament accumulates its mass when it forms.} normalized by the scale height ($R_0$), the plasma $\beta$\,\footnote{The ratio of thermal to magnetic pressure.} of the surrounding interstellar gas ($\beta_0$), and center-to-surface density ratio ($\rho_\mathrm{c} / \rho_\mathrm{s}$).
Figure \ref{fig:plummer_profile}(a) shows an example cross-sectional profile of one of his filament models.
Due to the axisymmetric mass accretion onto the filament during filament formation and evolution, the \emph{B}-field shows a pinched structure dragged toward the filament center \citep[see also, e.g.,][]{2021ApJ...911...15B}.

In Figures \ref{fig:plummer_profile}(c)--(h), we show the cross-section profiles of \emph{I} and PI when we observe the model filament shown in Figure \ref{fig:plummer_profile}(a) with various orientations.
To estimate the profiles, we assume a homogeneous dust alignment and dust properties in the filament.
This is to demonstrate that the $\mathrm{FWHM}(\mathrm{PI}) / \mathrm{FWHM}(I)$ variation can be caused solely by a pinched \emph{B}-field morphology without changing the dust alignment level and dust properties.
We also assume that the filament is optically thin for submillimeter radiation at 850 $\mu$m.
Under these assumptions, we can estimate \emph{I} and PI for each LOS looking through the filament as follows \citep{2015ApJ...807...47T}:
\begin{eqnarray}
I &=& \int \epsilon~\rho \left( 1 - P_\mathrm{max} \left( \frac{\cos^2 \gamma}{2} - \frac{1}{3} \right) \right) ds,\label{eq:I}\\
Q &=& P_\mathrm{max} \int \epsilon~\rho~\cos 2\psi~\cos^2 \gamma~ds,\\
U &=& P_\mathrm{max} \int \epsilon~\rho~\sin 2\psi~\cos^2 \gamma~ds,\\
PI &=& \left(Q^2+U^2 \right)^{1/2},\label{eq:PI}\\
P &=& PI/I,\label{eq:P}
\end{eqnarray}
where the integration is performed along the LOS;
$\epsilon$ is the dust emissivity;
$\rho$ is the gas density;
$P_\mathrm{max}$ is the maximum possible polarization fraction, for which we adopt a fiducial value of 0.15 (e.g., \citealp{2015ApJ...807...47T,2018MNRAS.474.5122K});
$\gamma$ is the offset angle of the local \emph{B}-field with respect to the POS (see Figure \ref{fig:plummer_profile}(b));
and $\psi$ is the local \emph{B}-field position angle.

If the \emph{B}-field is pinched in a filament, $\psi$ and $\gamma$ in Equations (\ref{eq:I})--(\ref{eq:PI}) vary spatially.
The $\psi$ variation along the LOS causes geometrical depolarization. 
The spatial variation of this geometrical depolarization and $\gamma$ as a function of angular distance from the filament's major axis result in different profiles of PI from that of \emph{I} (Figures \ref{fig:plummer_profile}(c)--(h)).
The greater the degree of \emph{B}-field pinch, the more significant the difference between the two profiles.

\begin{figure*}[tp]
\centering
\includegraphics[width=\linewidth]{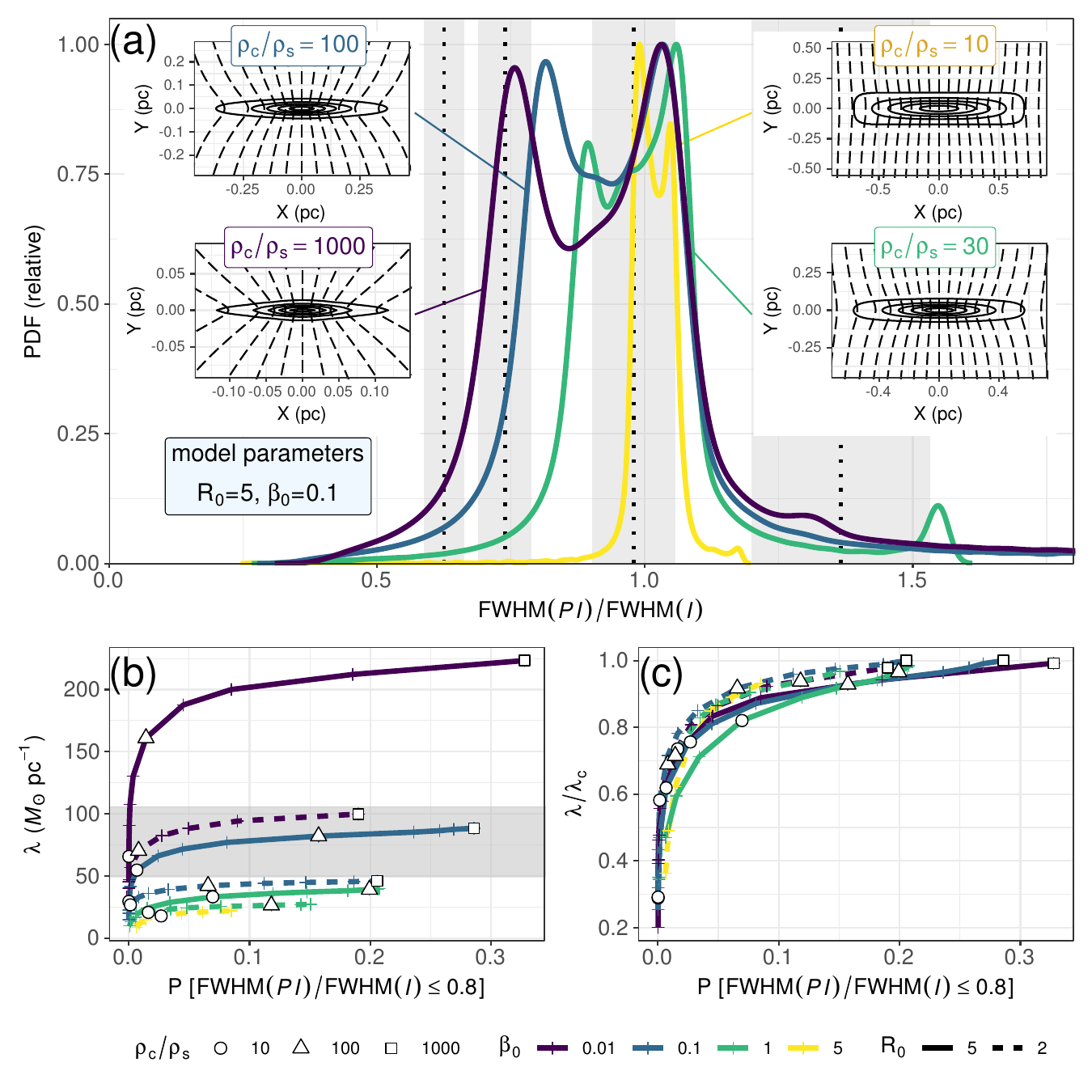}
\caption{
(a) PDF of the $\mathrm{FWHM}(\mathrm{PI})/\mathrm{FWHM}(I)$ ratio obtained for typical parameters of the model by \citet{2014ApJ...785...24T} when we observe the filament from a random orientation in 3D space (colored lines).
We assume $R_0 = 5$, $\beta_0=0.1$, and $\rho_c/\rho_s=10,~30,~100,$ and 1000 in the figure.
See text for the definition of $R_0,~\beta_0$, and $\rho_c/\rho_s$.
Vertical dotted lines are the observed ratios (beam-deconvolved) with their $\pm 1 \sigma$ error indicated as shaded areas (segments B, A, C, and D from left to right; see Table \ref{table:FWHM}).
Insets show cross-sectional views of the filament gas density and \emph{B}-field structure corresponding to each parameter set.
Solid contours are relative density $\rho/\rho_c=0.1,~0.3,~\ldots,0.9$, and dashed lines represent \emph{B}-field lines.
We apply the same assumption as in Figures \ref{fig:plummer_profile}(c)--(h) when converting the radial distribution of the gas to a physical scale in parsecs.
(b) Comparison of the filament line mass ($\lambda$) and the probability that we observe $\mathrm{FWHM}(\mathrm{PI})/\mathrm{FWHM}(I) \leqslant 0.8$ for each model parameter set.
We assume a gas temperature of 15 K and a corresponding critical line mass of thermally supported filaments of
$\lambda_\mathrm{c,~thermal} = 2c_s^2/G \simeq 24~M_\odot~\mathrm{pc^{-1}}$ \citep{1963AcA....13...30S,1964ApJ...140.1056O,1997ApJ...480..681I}.
The color of the lines indicates $\beta_0 = 0.01$, 0.1, 1, and 5, and the line types (solid and dashed) indicate $R_0 = 5$ and 2, respectively.
Tick marks in the colored lines indicate $\rho_c/\rho_s = $2, 3, 5, 10, 20, 30, 50, 100, 200, 300, 500, and 1000.
The larger $\rho_c/\rho_s$ data may not have been plotted due to numerical difficulties in obtaining equilibrium solutions at such high densities.
The shaded area indicates the estimated line mass of the filament that contains regions A--D (49--106 $M_\odot~\mathrm{pc}^{-1}$; \citetalias{2020ApJ...899...28D}).
(c) Same as panel (b) but for the ratio of $\lambda$ to the magnetically critical value ($\lambda/\lambda_c$).
}
\label{fig:plummer_p_distribution_plot_rot}
\end{figure*}

We show in Figure \ref{fig:plummer_p_distribution_plot_rot}(a) the probability distribution function (PDF) of $\mathrm{FWHM}(\mathrm{PI}) / \mathrm{FWHM}(I)$ when we observe a filament with a given parameter set from random orientations in the 3D space\footnote{We estimate the probability distribution of $\mathrm{FWHM}(\mathrm{PI}) / \mathrm{FWHM}(I)$ by randomly changing $\theta$ (the relative angle of the filament with respect to the LOS) and $\phi$ (the rotation angle of the magnetic field with respect to the long axis of the filament) 6208 times for each parameter set. See Figure \ref{fig:plummer_profile}(b) for definitions of $\theta$ and $\phi$.}, predicted by the Tomisaka model for typical sets of parameters.
Insets in Figure \ref{fig:plummer_p_distribution_plot_rot}(a) show cross-sectional views of the filament gas density and \emph{B}-field structure corresponding to each parameter set.
When $\rho_c/\rho_s$ is small and the \emph{B}-field pinching inside the filament is negligible (e.g., the case of $\rho_c/\rho_s = 10$ in the figure), the probability of the $\mathrm{FWHM}(\mathrm{PI}) / \mathrm{FWHM}(I)$ ratio concentrates around 1.
On the other hand, when $\rho_c/\rho_s$ becomes large, the \emph{B}-field is pinched significantly toward the filament center to support the filament gas against its self-gravity (e.g., the case of $\rho_c/\rho_s = 1000$ in the figure).
This significant pinch of the \emph{B}-field results in a significant probability of $\mathrm{FWHM}(\mathrm{PI}) / \mathrm{FWHM}(I) \ne 1$, which is consistent with the observed results.

Indeed, we find $\mathrm{FWHM}(\mathrm{PI}) / \mathrm{FWHM}(I) \leqslant 0.8$ for two of the four observations (Section \ref{sec: results}).
It is difficult to reproduce $\mathrm{FWHM}(\mathrm{PI}) / \mathrm{FWHM}(I) < 1$ by modifying the dust alignment level,  since the level needs to increase inside the dense filament to produce $\mathrm{FWHM}(\mathrm{PI}) / \mathrm{FWHM}(I) < 1$ (Section \ref{sec: results}).

To verify whether the geometrical depolarization predicted by the Tomisaka model can reproduce the observation, we estimate the probability that the model predicts $\mathrm{FWHM}(\mathrm{PI}) / \mathrm{FWHM}(I) \leqslant 0.8$.
We plot the probability as a function of filament line mass ($\lambda$) in Figure \ref{fig:plummer_p_distribution_plot_rot}(b), and $\lambda$'s ratio to the magnetically critical line mass ($\lambda/\lambda_c$) in Figure \ref{fig:plummer_p_distribution_plot_rot}(c).

We find that the model predicts up to $\simeq 30$\% probability of reproducing $\mathrm{FWHM}(\mathrm{PI}) / \mathrm{FWHM}(I) \leqslant 0.8$ depending on the choice of input parameters.
\citetalias{2020ApJ...899...28D} estimated $\lambda \sim 50$--$100~ M_\odot \mathrm{pc}^{-1}$ for the filaments that contain segments A--D (shaded area in Figure \ref{fig:plummer_p_distribution_plot_rot}(b)).
Figure \ref{fig:plummer_p_distribution_plot_rot}(b) suggests that the cases of $R_0=2,~\beta_0=0.01$--$0.1$ or $R_0=5,~\beta_0=0.1$--$1$ are preferred.
In addition to that, Figure \ref{fig:plummer_p_distribution_plot_rot}(c) suggests that the near-critical (or supercritical) values of $\lambda$, with $\lambda / \lambda_\mathrm{c} > 0.9$, are preferred to reproduce the observation using the model.
Since the \emph{B}-field pinching depends on $\lambda / \lambda_\mathrm{c}$, the larger the $\lambda / \lambda_\mathrm{c}$, the more significant the amount of \emph{B}-field pinching and, as a result, the larger the probability of $\mathrm{FWHM}(\mathrm{PI}) / \mathrm{FWHM}(I) \leqslant 0.8$.

Thus, we conclude that the observed variation of the $\mathrm{FWHM}(\mathrm{PI}) / \mathrm{FWHM}(I)$ ratio can be explained by pinched \emph{B}-field structure inside the filament.
We note that the data with $\mathrm{FWHM}(\mathrm{PI}) / \mathrm{FWHM}(I) < 1$ are obtained from a single filament (Figure \ref{fig:filament_section_pos}).
Therefore, the observed PDF may be biased toward small values of $\mathrm{FWHM}(\mathrm{PI}) / \mathrm{FWHM}(I)$.
We need more observations to discuss further details.
At this moment, however, we can point out that the PDF estimated by the model should predict $\mathrm{FWHM}(\mathrm{PI}) / \mathrm{FWHM}(I) \ne 1$ with sufficient probability.
In other words, our observations strongly suggest a significant pinch of the \emph{B}-field inside the filament.

\section{Discussion}
\label{sec:discussion}

In the previous section, we have shown that the Tomisaka model can naturally explain the variation of observed $\mathrm{FWHM}(\mathrm{PI}) / \mathrm{FWHM}(I)$ only by changing the filament's viewing angle.
It does not mean, however, that we can exclude other possibilities.

When assessing the model predictions, we assume that the dust alignment level and dust properties are homogeneous inside and outside the filament.
We think it is unlikely that the dust alignment level increases at the filament center to produce $\mathrm{FWHM}(\mathrm{PI}) / \mathrm{FWHM}(I) < 1$ (see Section \ref{sec: results}).
On the other hand, it may be possible that the spatial variation of the dust properties causes more efficient polarization of emission at the filament center.
However, in such a case, it would be necessary to explain naturally how \emph{P}, which generally decreases in the molecular cloud, becomes locally higher in, e.g., segment B.

In contrast to the quiescent filament interior, the dust alignment level and dust properties may be spatially nonuniform in active regions.
We show two example emission profiles at SVS 3 and SVS 13A in Figure \ref{fig:filament_section_profile_app} (see Appendix \ref{sec:active regions} for the derivation details).
\begin{figure}[tp]
\centering
\includegraphics[width=1\linewidth]{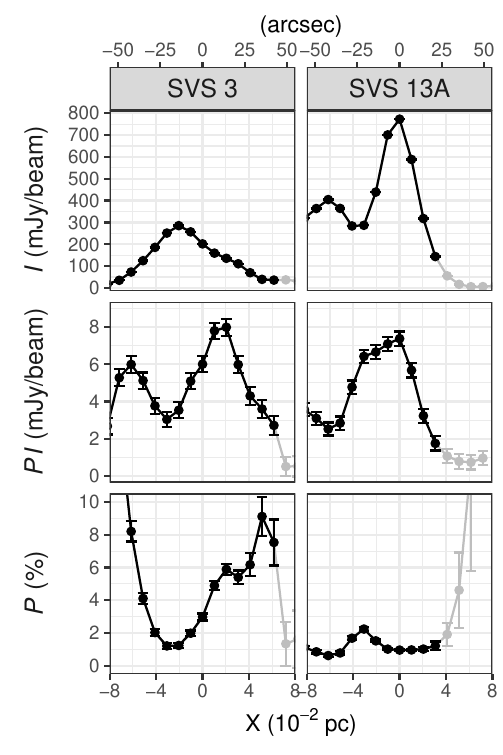}
\caption{
Same as Figure \ref{fig:filament_section_profile} but for the filaments' cross-sectional profiles in two active regions (SVS 3 and SVS 13A).
}
\label{fig:filament_section_profile_app}
\end{figure}
Both the \emph{I} and PI profiles show a distorted shape and do not match with each other, which is considerably different from the quiescent filaments seen in regions A--D, where a single Gaussian fits well each profile.
Furthermore, the $\mathrm{PI} / \delta \mathrm{PI}$ value in these regions shows considerable spatial variation, and our criterion of achieving $\mathrm{PI}/\delta \mathrm{PI} \geqslant 3$ over more than three consecutive beams is not met.
This distortion of emission profiles is potentially due to the spatial variation of the dust properties and alignment level.
We limit ourselves to discussing \emph{B}-field structure in quiescent filaments and do not discuss these profiles in active regions further in this paper.

The Tomisaka model assumes that an initially uniform \emph{B}-field intersects a cylindrical filament perpendicularly.
However, it may be possible to consider a more complex \emph{B}-field.
\citet[][also see \citealp{2018MNRAS.481.2507R}]{2000ApJ...544..830F} showed that a helical \emph{B}-field structure can reproduce observed decreases in \emph{P} in dense molecular cores (``polarization holes'').
We do not discuss the comparison between helical \emph{B}-fields and our observations in this paper.
Instead, we want to emphasize that the lateral \emph{B}-field assumed in the Tomisaka model explains well the observed offset angle between the filament and the \emph{B}-field (\citetalias{2020ApJ...899...28D}) and can also reproduce the polarization hole as demonstrated in Figure \ref{fig:plummer_profile}(h), where the \emph{B}-field at the center of the filament is almost parallel to the LOS.

Disturbance of the \emph{B}-field due to turbulence may be less in the center of quiescent filaments.
In such a case, geometrical depolarization due to unresolved small-scale \emph{B}-field turbulence may be less inside the filament than outside, resulting in a larger \emph{P} and $\mathrm{FWHM}(\mathrm{PI}) / \mathrm{FWHM}(I) < 1$.
However, if it is the primary cause of $\mathrm{FWHM}(\mathrm{PI}) / \mathrm{FWHM}(I)$ variation, we need to understand why \emph{B}-field turbulence is low inside one quiescent filament and high in another quiescent filament where $\mathrm{FWHM}(\mathrm{PI}) / \mathrm{FWHM}(I) > 1$ and the central \emph{P} is small.

The variation of \emph{P} inside the filament may not only be due to the geometrical effect of the pinched \emph{B}-field.
In fact, the observed \emph{P} is $\sim 1/3$ of the model-predicted values assuming the same efficiency of polarized dust emission in the diffuse region and the filament interior (Figure \ref{fig:plummer_profile}).
Thus, the efficiency of polarized dust emission may be reduced inside the filament in addition to the geometrical depolarization.

We note that the filaments in NGC 1333 may not be in exact magnetohydrostatic equilibrium, as suggested by \citet{2017A&A...606A.123H}.
However, the equilibrium solutions of \citet{2014ApJ...785...24T} are probably a good representation of the structures that arise in the dynamical formation of a filament, even if equilibrium has not been achieved.
In fact, Figure \ref{fig:plummer_p_distribution_plot_rot}(c) indicates that the line mass may be supercritical ($\lambda / \lambda_\mathrm{c} > 0.9$).
If the filament is supercritical and the gas in the filament, which is dynamically contracting, is dragging the \emph{B}-field lines toward its center, the observations should be well reproduced.

The Tomisaka model predicts a flattened filament profile, as shown in Figures \ref{fig:plummer_profile}(a) and \ref{fig:plummer_p_distribution_plot_rot}(a).
It results in a variation in the FWHM of the filaments' \emph{I} profiles observed from different directions, as demonstrated in Figures \ref{fig:plummer_profile}(c)--(h).
This variation due to the viewing angle could be an explanation of the variance on the canonical 0.1 pc FWHM reported by, e.g., \citet[][also see \citealp{2016ApJ...831...46A}]{2011A&A...529L...6A,2019A&A...621A..42A}.

Following the discussion above, we conclude that the pinched \emph{B}-field predicted by the Tomisaka model is a plausible explanation of the observed $\mathrm{FWHM}(\mathrm{PI}) / \mathrm{FWHM}(I)$ variation of quiescent filaments in NGC 1333.
The \emph{B}-field that is dragged inward by the contracting ISM has been observed in some cases as an ``hourglass'' structure of the \emph{B}-field around star-forming cores and is recognized as a sign of magnetically regulated collapse of spherical cores \citep[e.g.,][]{2006Sci...313..812G,2013ApJ...769L..15S,2017ApJ...848..110K,2018MNRAS.477.2760M,2019ApJ...879...25K,2020ApJ...892..152H}.
We do not see such an hourglass morphology in the observed \emph{B}-field structure (Figure \ref{fig:filament_section_pos}; however, see \citealp{2017ApJ...846..122P} for an example found in the Orion A filament).
Instead, our result is an indirect observation of the possibly pinched internal \emph{B}-field structure in dense interstellar filaments in star-forming regions as a result of an axisymmetrical contraction of the filament ISM.
The $\mathrm{FWHM}(\mathrm{PI}) / \mathrm{FWHM}(I)$ ratio, especially its deviation from 1 as we observed, may be an essential indicator of the degree of \emph{B}-field pinching in filaments.

\section{Conclusions}\label{conclusions}

\label{ref:conclusions}

We performed submillimeter-wavelength polarization observations using SCUBA-2/POL-2 on JCMT and characterized the POS magnetic field (\emph{B}-field) of NGC 1333.
Following \citetalias{2020ApJ...899...28D}, we investigated the 3D \emph{B}-field distribution inside filaments that do not show evident star-forming activity and thus are thought to retain their initial formation state.
We found that the filaments' FWHMs in PI are not the same as those in \emph{I}, with two segments being appreciably narrower in PI ($\mathrm{FWHM}(\mathrm{PI})/\mathrm{FWHM}(I) \simeq 0.7$--0.8) and one segment being wider ($\mathrm{FWHM}(\mathrm{PI})/\mathrm{FWHM}(I) \simeq 1.3$) out of four investigated filament segments.
None of the profiles of \emph{P} inside filaments show an anticorrelation with \emph{I} normally found in molecular clouds and protostellar cores.

We showed that the magnetohydrostatic equilibrium solution of a filament threaded by a lateral magnetic field \citep{2014ApJ...785...24T} well reproduces the observed variation of $\mathrm{FWHM}(\mathrm{PI}) / \mathrm{FWHM}(I)$, although we do not exclude other possibilities.
The \emph{B}-field inside a filament is radially dragged inward (pinched) along with the matter contraction during the formation of the filament.
This pinched \emph{B}-field structure causes the local directional changes of the \emph{B}-field within the filament and thereby the geometrical depolarization that can reproduce local variations of PI and \emph{P}.
The appreciable deviation of $\mathrm{FWHM}(\mathrm{PI}) / \mathrm{FWHM}(I)$ from 1 indicates that the \emph{B}-field is pinched significantly.
In other words, quiescent filaments in NGC 1333 are suggested to be magnetically near-critical or supercritical.

The $\mathrm{FWHM}(\mathrm{PI}) / \mathrm{FWHM}(I)$ ratio can provide important information about the pinched \emph{B}-field structure expected inside the filament and, consequently, help us better understand the role of the \emph{B}-field in the formation of filaments.

%\newpage

\acknowledgments

The authors thank the anonymous referee for the valuable comments.
The JCMT is operated by the EAO on behalf of NAOJ, ASIAA, KASI, and CAMS, as well as the National Key R\&D Program of China (No. 2017YFA0402700).
Additional funding support is provided by the STFC and participating universities in the UK and Canada.
This research used the facilities of the Canadian Astronomy Data Centre operated by the National Research Council of Canada with the support of the Canadian Space Agency.
This research has also made use of the simbad database and of NASA's Astrophysics Data System Bibliographic Services.
Part of the data analysis was carried out on the open use data analysis computer system at the Astronomy Data Center (ADC) of the National Astronomical Observatory of Japan.
This research has been supported by Grants-in-Aid for Scientific Research (25247016, 18H01250, 19H0193, 19K03919) from the Japan Society for the Promotion of Science and a crowdfunding run by academist project No. 25 (\href{https://academist-cf.com/}{https://academist-cf.com/projects/25}).
M.M. is supported by JSPS KAKENHI grant No. 20K03276.
C.L.H.H. acknowledges the support of the NAOJ Fellowship and JSPS KAKENHI grants 18K13586 and 20K14527.
D.J. and J.D.F. are supported by the National Research Council of Canada and by Natural Sciences and Engineering Research Council of Canada (NSERC) Discovery Grants.
W.K. was supported by the New Faculty Startup Fund from Seoul National University.
L.F. acknowledges the support of the Ministry of Science and Technology of Taiwan under grant MOST107-2119-M-001-031-MY3 and Academia Sinica under grant AS-IA-106-M03.
K.Q. is partially supported by National Key R\&D Program of China No. 2017YFA0402600, and acknowledges the National Natural Science Foundation of China (NSFC) grant U1731237.

%% To help institutions obtain information on the effectiveness of their 
%% telescopes the AAS Journals has created a group of keywords for telescope 
%% facilities.
%
%% Following the acknowledgments section, use the following syntax and the
%% \facility{} or \facilities{} macros to list the keywords of facilities used 
%% in the research for the paper.  Each keyword is check against the master 
%% list during copy editing.  Individual instruments can be provided in 
%% parentheses, after the keyword, but they are not verified.

%\vspace{5mm}
\facilities{JCMT (SCUBA-2/POL-2)}

%% Similar to \facility{}, there is the optional \software command to allow 
%% authors a place to specify which programs were used during the creation of 
%% the manuscript. Authors should list each code and include either a
%% citation or url to the code inside ()s when available.

\software{Starlink \citep{2014ASPC..485..391C}, astropy \citep{2013A&A...558A..33A}}

%% Appendix material should be preceded with a single \appendix command.
%% There should be a \section command for each appendix. Mark appendix
%% subsections with the same markup you use in the main body of the paper.

%% Each Appendix (indicated with \section) will be lettered A, B, C, etc.
%% The equation counter will reset when it encounters the \appendix
%% command and will number appendix equations (A1), (A2), etc. The
%% Figure and Table counter will not reset.

%\vspace*{5mm}
%\newpage

\appendix
\setcounter{table}{0}
\renewcommand{\thetable}{A.\arabic{table}}
\setcounter{figure}{0}
\renewcommand{\thefigure}{A.\arabic{figure}}

\section{Evaluation of Stokes Parameters and Their Sensitivity as a Function of Spatial Frequency}
\label{sec:polynomial}

We estimate the Stokes \emph{I}, \emph{Q}, and \emph{U} parameters at each position in the sky from the standard pipeline $pol2map$ \citep[][software version on 2018 November 17]{Parsons2018} in {\sc Starlink} (\citealp {2014ASPC..485..391C}; see \citetalias{2020ApJ...899...28D} for further details).
As described in \citetalias{2020ApJ...899...28D}, we apply a 2D least-squares fit of a second-order polynomial using a Gaussian kernel to the Stokes \emph{I}, \emph{Q}, and \emph{U} data.
We debias PI to correct a possible offset due to the square-sum of the \emph{Q} and \emph{U} errors.
We then use only $\mathrm{PI}/\delta \mathrm{PI} \geqslant 3$ data for the following analyses to make the debiasing effect negligible.
The assumed FWHM of the Gaussian kernel is $14\arcsec.1$, which corresponds to the JCMT/POL-2 beam at $850~\mu$m if we fit the beam with a single Gaussian profile \citep{2013MNRAS.430.2534D}.

\begin{figure*}[tp]
\centering
\includegraphics[width=0.5\linewidth]{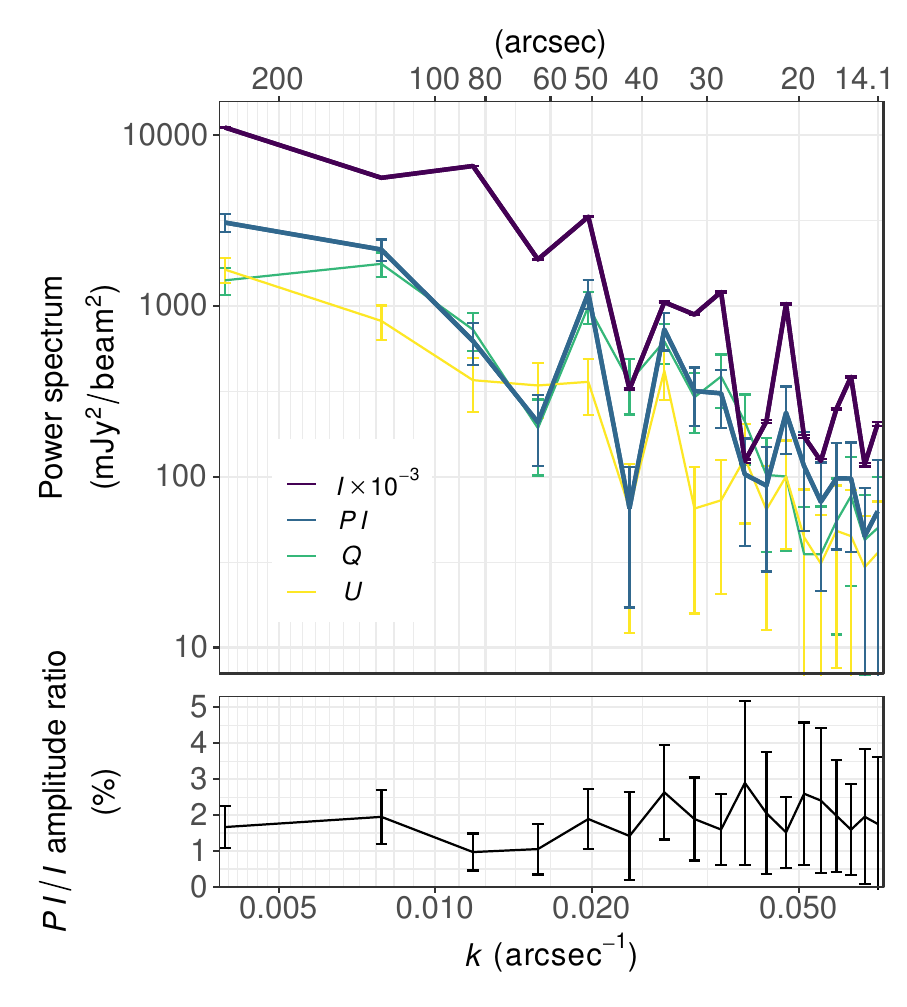}
\caption{
Top panel: spatial power spectra of \emph{I}, \emph{Q}, \emph{U}, and PI signals measured along the blue dashed line in Figures \ref{fig:filament_section_pos} and \ref{fig:N1333_Pezzuto_temp}.
Bottom panel: ratio between PI and \emph{I} spectrum amplitudes.
}
\label{fig:filament_FFT}
\end{figure*}

The Stokes \emph{I} intensity is estimated by removing background atmospheric emission from the observed data, resulting in a reduced sensitivity for the large-scale emission.
On the other hand, \emph{Q} and \emph{U} are measured in AC mode performed while rotating a half-wave plat; thus, the sensitivity is retained for the large-scale emission.
We estimate the spatial power spectra of these emissions and PI along the blue dashed line in Figure \ref{fig:filament_section_pos} to test if the sensitivities of \emph{I}, \emph{Q}, and \emph{U} are consistent for the small-scale emission, which is the subject of this paper.
As shown in Figure \ref{fig:filament_FFT} (top panel), the spectral shapes are consistent with each other.
Also, we find the nearly constant \emph{PI/I} ratio of spectral amplitude, as shown in Figure \ref{fig:filament_FFT} (bottom panel).
Thus, we confirm that the sensitivities of \emph{I} and PI are consistent for the small-scale $( \leqslant 2\arcmin)$ structures.

\section{Evaluation of Filaments' Cross-sectional Profiles}
\label{sec:profiles}

\begin{figure*}[tp]
\centering
\includegraphics[width=\linewidth]{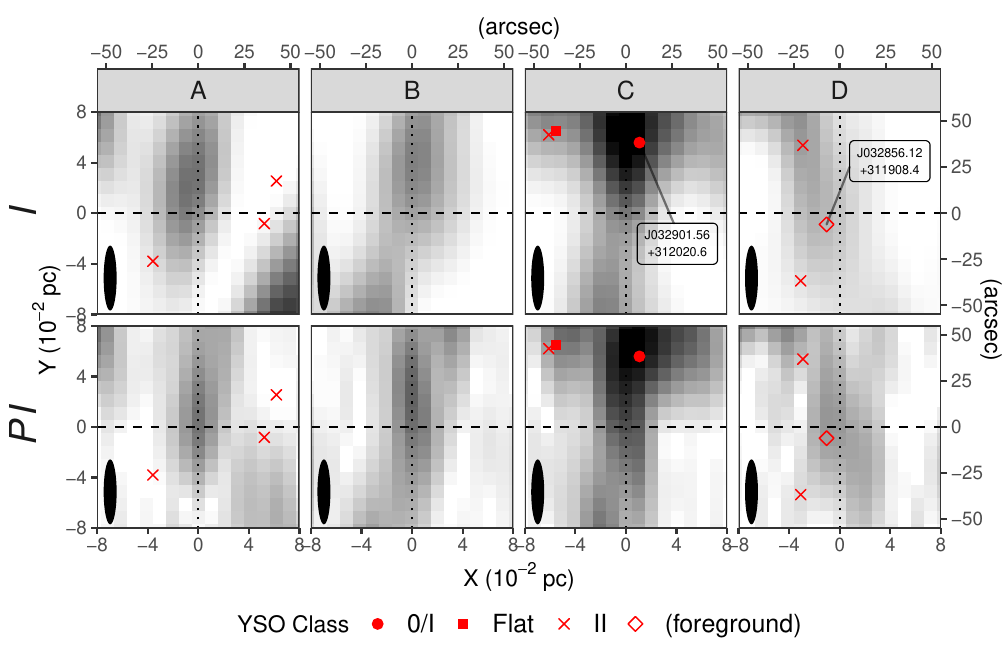}
\caption{Spatially smoothed distribution of $850~\mu\mathrm{m}$ Stokes $I$ and polarized emission $PI$ of the four selected regions.
The positions of the regions are indicted in Figure \ref{fig:filament_section_pos}.
We assume $14\arcsec.1$ spatial resolution (FWHM) across the filament direction and $70\arcsec.5$ in the filament direction, to achieve good SNRs in the cross-sectional profiles of filaments.
YSOs found near the positions of cross-sections \citep{2009ApJS..181..321E,2015AJ....150...40Y} are marked with red symbols.
We set the position of segment C so that we can exclude one Class 0/I YSO (J032901.56+312020.6) from the evaluation of the cross-section (Section \ref{sec:ObsReduce}).
We ignore one Class II YSO (J032856.12+311908.4) in region D as a foreground source since the estimated $A_V = 0$ mag \citep{2009ApJS..181..321E}.
The assumed Gaussian beams are indicated in the bottom left corners of the figures.
}
\label{fig:filament_section_map}
\end{figure*}

We select the data with $\mathrm{PI}/\delta \mathrm{PI} \geqslant 5$ for segments A, B, and C and $\mathrm{PI}/\delta \mathrm{PI} \geqslant 3$ for segment D and estimate the position angles of the major axes of the filaments at each position.
By assuming that we can approximate filaments locally by straight lines, we estimate the position angles by least-squares linear fits to the spatial structures in \emph{I}, weighted by \emph{I} intensity.
The estimated position angles are listed in Table \ref{tab:PA}, together with those of active regions described in Appendix \ref{sec:active regions}.
We then set the FWHM of the Gaussian kernel as $14\arcsec.1$ in the direction perpendicular to the filament and $70\arcsec.5$, or five times the beam, in the direction along the filament and reestimate the Stokes \emph{I}, \emph{Q}, and \emph{U} parameters at each position in the sky to estimate the cross-sectional profiles of filaments with improved S/Ns.
We show the estimated Stokes \emph{I} maps and the PI maps, which are estimated from the Stokes \emph{Q} and \emph{U} maps in Figure \ref{fig:filament_section_map}, together with the assumed beam profile.

\begin{table}
    \centering
    \caption{Estimated Position Angles of the Filaments}
    \begin{tabular}{lr}
    \hline
    \hline
    \multicolumn{1}{c}{Segment} & \multicolumn{1}{c}{Position Angle}\\
    & \multicolumn{1}{c}{(deg)}\\
    \hline
    A & $-20.9 \pm 11.6$\\
    B & $-7.8 \pm \ \,4.4$\\
    C & $-14.4 \pm \ \,3.7$\\
    D & $12.5 \pm 12.9$\\
    SVS 3 & $-35.9 \pm \ \,1.8$\\
    SVS 13A & $-47.1 \pm \ \,3.5$\\
    \hline
    \multicolumn{2}{p{0.45\textwidth}}{{\bf Note.} See text for the details of the derivation.}
    \end{tabular}
    \label{tab:PA}
\end{table}

We estimate errors by performing a Monte Carlo simulation by assuming a Gaussian distribution for the \emph{pol2map}-estimated error.
We repeat the polynomial fitting 1000 times with Gaussian random errors at each position.
We take the mean of 1000 samples as the estimated values and the standard deviation as the estimation error of each position.

We fit the estimated cross-sectional profiles of \emph{I} and PI with Gaussian curves as shown in Figure \ref{fig:filament_section_profile}.
We use the data that satisfy both $I / \delta I \geqslant 3$ and $\mathrm{PI} / \delta \mathrm{PI} \geqslant 3$ for fitting \emph{I} and PI profiles.
Practically, all data points with $\mathrm{PI} / \delta \mathrm{PI} \geqslant 3$ satisfy $I / \delta I \geqslant 3$.
When making Gaussian fittings, we assume the baseline intensity of \emph{I} and PI as zero and judge the residual low S/N signal seen in PI (Figure \ref{fig:filament_section_profile}) as residual noise of the debiasing.
If we estimate the Gaussian baseline level with these signals, positive baseline levels lead to narrow PI profiles.

We then deconvolve the fitted Gaussian by the observational beam by assuming the beam as a $14.\arcsec 1$ single Gaussian \citep{2013MNRAS.430.2534D} as follows:
\begin{equation}
    \sigma_\mathrm{deconvolved} = \sqrt{\sigma_\mathrm{observed}^2 - \sigma_\mathrm{beam}^2},
\end{equation}
where $\sigma_\mathrm{deconvolved}$,  $\sigma_\mathrm{observed}$, and $\sigma_\mathrm{beam}$ are the variances of deconvolved, observed, and the beam's Gaussian, respectively.

\section{Dust Temperature}
\label{sec:dust temperature}

\begin{figure*}[tp]
\centering
\includegraphics[width=\linewidth]{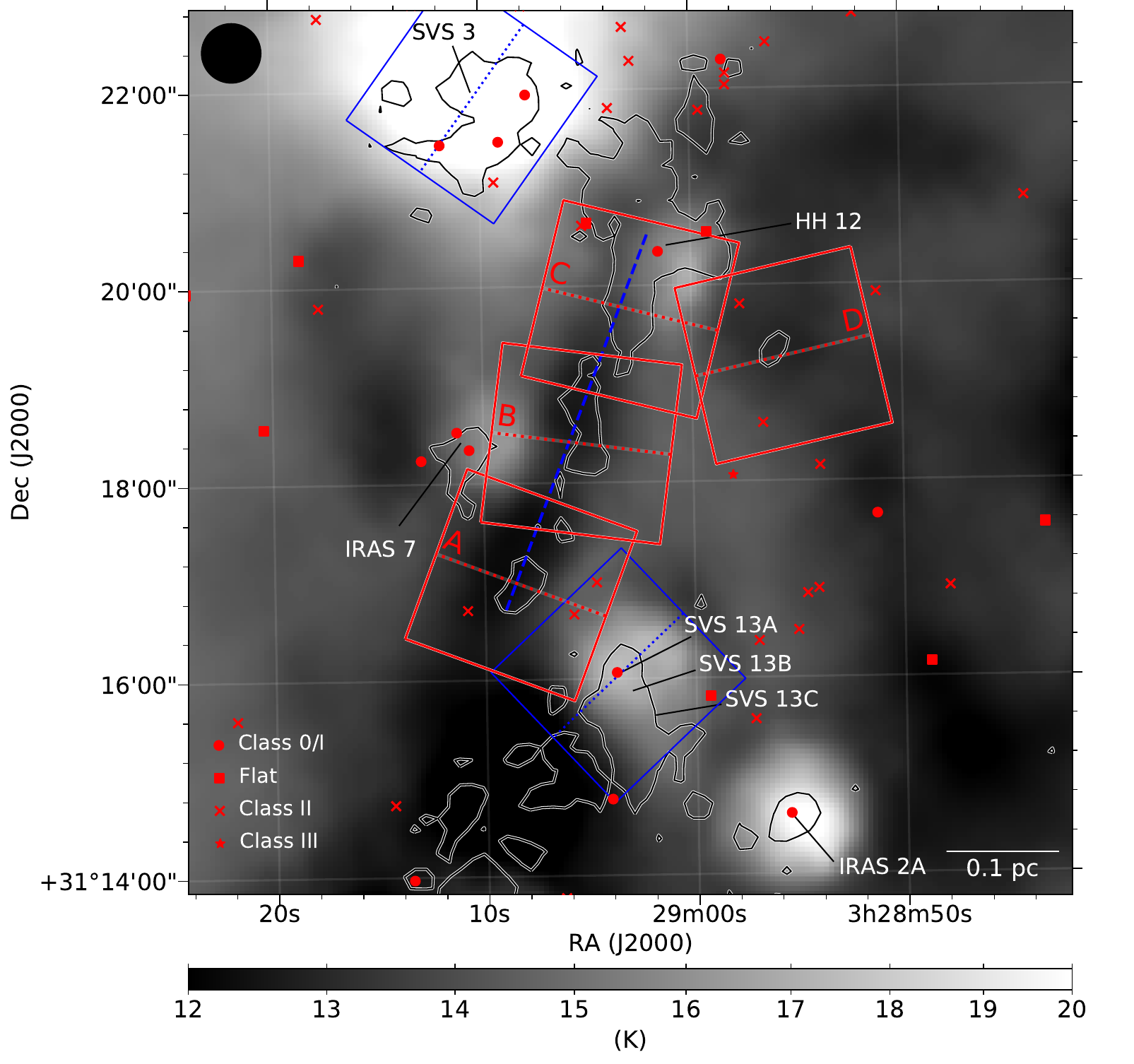}
\caption{
Same as Figure \ref{fig:filament_section_pos} but for dust color temperature estimated by Herschel and Planck observations \citep{2021A&A...645A..55P}.
The resolution of the map ($36\arcsec.1$) is shown in the upper left corner of the figure.
}
\label{fig:N1333_Pezzuto_temp}
\end{figure*}

We refer to \citet[][Figure \ref{fig:N1333_Pezzuto_temp}]{2021A&A...645A..55P} and estimate the dust color temperature at the filament profiles.
We confirm that there are no significant heating sources embedded in the regions, and there is no significant nonuniform external heating from nearby sources.
The estimated dust temperature at each profile is $T=12.2$--15.8 K for region A, 12.6--16.0 K for region B, 13.1--17.0 K for region C, and 12.9--13.7 K for region D.

\section{Filament Profiles in Active Regions}
\label{sec:active regions}

\begin{figure*}[tp]
\centering
\includegraphics[width=0.5\linewidth]{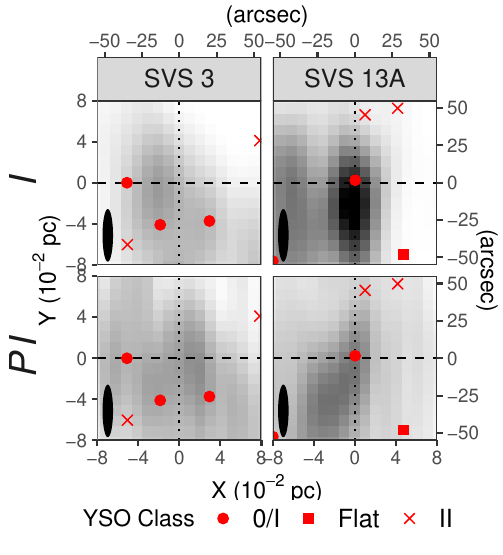}
\caption{
Same as Figure \ref{fig:filament_section_map} but for maps of two active regions.
The two regions are centered at SVS 3 and SVS 13A.
}
\label{fig:filament_section_map_app}
\end{figure*}

We estimate two example cross-sectional profiles of filaments in the active region.
We set the positions at SVS 3 and SVS 13A, as shown in Figures \ref{fig:filament_section_pos} and \ref{fig:N1333_Pezzuto_temp}.
For these regions, we select the data with $I \geqslant 25$ (mJy beam$^{-1}$) and estimate the local filament position angles as described in Appendix \ref{sec:profiles}.
We do not restrict the data based on $\mathrm{PI}/\delta \mathrm{PI}$ to estimate the filament's position angles, as the S/N of PI largely fluctuates in the active regions.
The estimated position angles are shown in Table \ref{tab:PA}.
We then estimate \emph{I} and PI intensity maps smoothed along the major axes of the filaments as described in Appendix \ref{sec:profiles}.
The estimated intensity maps are shown in Figure \ref{fig:filament_section_map_app}.
The cross-sectional profiles of \emph{I}, PI, and \emph{P} are shown in Figure \ref{fig:filament_section_profile_app}.

%% For this sample we use BibTeX plus aasjournals.bst to generate the
%% the bibliography. The sample63.bib file was populated from ADS. To
%% get the citations to show in the compiled file do the following:
%%
%% pdflatex sample63.tex
%% bibtext sample63
%% pdflatex sample63.tex
%% pdflatex sample63.tex

\bibliography{paper2}{}
\bibliographystyle{aasjournal}

%% This command is needed to show the entire author+affiliation list when
%% the collaboration and author truncation commands are used.  It has to
%% go at the end of the manuscript.
%\allauthors

%% Include this line if you are using the \added, \replaced, \deleted
%% commands to see a summary list of all changes at the end of the article.
%\listofchanges

\end{document}